\documentstyle[12pt]{article}
\def\fracl#1#2{\frac{\displaystyle #1}{\displaystyle #2}}

\begin{document}

\begin{center}
{\sc V.Bashkov, M.Malakhaltsev}\\
{\large\bf
Geometry of rotating disk and the Sagnac effect}
\end{center}

In this paper we demonstrate that  subsequent 
application of Lorentz transformations to  the cylindrical coordinates on a rotating disk leaves 
the form $ds^2 = c^2 dt^2 - dr^2 - r^2 d\phi^2 - dz^2$ invariant. Therefore, the geometry on  
rotating disk is the Euclidean geometry, and any experiment which do not involve 
tidal forces or Coriolis forces  cannot identify either the disk rotates or not. 
We also show that, from the point of view of external inertial observer, 
the difference in the transit times for the light running along a circle of radius $R$ 
in the opposite directions (with respect to the rotation) is 
$\Delta t = 2 \fracl{\omega}{c^2}S$,
where $S$ is the area the circle.

We start with a historical review. In [2]  H.A.Lorentz 
considered transformation from a resting coordinate system  
to another coordinate system moving with respect to the first one with velocity $v$ 
as composition of the following transformations (see [1]):
\par\noindent
1) The Galilean transformation
\begin{equation} \label{E1}
x' = x - vt,  \quad y' = y, \quad z' = z, \quad t'=t.
\end{equation}
2) The transformation which gives the length shrinking with respect to  the moving coordinate
system and introduces the local time at any point of  moving coordinate system
\begin{equation} \label{E2}
x'' = \fracl{x'}{\sqrt{1 - \fracl{v^2}{c^2}}}, \quad
t'' = \sqrt{1 - \fracl{v^2}{c^2}} t' - \fracl{\fracl{v}{c^2}x' }{\sqrt{1 - \fracl{v^2}{c^2}}},
\end{equation}
where  $c$ is the light velocity.
\par
In [3] A.Poincare gave the formulas for the resulting transformation 
\begin{equation} \label{E3}
x'' = \fracl{x-vt}{\sqrt{1- \fracl{v^2}{c^2}}}; \quad y'' = y, \quad z'' = z,
\quad t'' = \fracl{t-\fracl{v}{c^2}x }{\sqrt{1- \fracl{v^2}{c^2}}}
\end{equation}
In what follows we shall use the infinitesimal versions of transformations  
(\ref{E1} -- \ref{E3}):
\par\noindent
$
\begin{array}{llll}
dx' = dx - v dt,   &dy' = dy,  &dz' = dz,   & dt' = dt; 
\\
dx'' =\fracl{dx'}{\sqrt{1- \fracl{v^2}{c^2}}}  & dy'' = dy',  &dz'' = dz' ;
& dt'' = \sqrt{1- \fracl{v^2}{c^2}}dt' - \fracl{\fracl{v}{c^2}dx }{\sqrt{1- \fracl{v^2}{c^2}}}
\\
dx'' = \fracl{dx-vdt}{\sqrt{1- \fracl{v^2}{c^2}}}
& dy' = dy,  & dz' = dz, 
& dt'' =  \fracl{dt-\fracl{v}{c^2}dx }{\sqrt{1- \fracl{v^2}{c^2}}}
\end{array}
$
\hfill
\parbox{1truecm}{
\begin{tabular}{r}
(\ref{E1}')\\
(\ref{E2}')\\
(\ref{E3}')
\end{tabular}}

It is important to note that the transformation of local time (\ref{E2}') consists of two 
transformations
\begin{equation} 
dt''_1 = \sqrt{1- \fracl{v^2}{c^2}}dt' \qquad
dt''_2 = \fracl{-\fracl{v}{c^2}dx }{\sqrt{1- \fracl{v^2}{c^2}}}
\end{equation}

Now let us consider coordinate transformations with respect to a rotating coordinate
system. Following M\"oller [4],
we intoduce an inertial reference system with cylindrical
coordinates ($r$, $\theta$, $z$,$t$), and a rotating coordinate system $S$ with coordinates 
($r'$, $\theta'$, $z'$,$t'$) which are expressed through ($r$, $\theta$, $z$,$t$) via the 
Galilean transformation (see (\ref{E1})):
\begin{equation} 
r' = r, \quad \theta' = \theta - \omega t, \quad z' = z,  \quad t' = t.
\end{equation}
where $\omega$ is the angle velocity of the disk.
Then the infinitesimal transformation takes the form
$$ 
dr' = dr, \quad d\theta' = d\theta - \omega dt, \quad dz' = dz,  \quad dt' = dt,
\eqno{(5')}
$$ 
Here $\omega$ is the angle velocity of the disk measured by the inertial observer.

Now, following H.A.Lorentz [2] we introduce new local coordinates via the
transformation
$$
dr'' = dr', \quad r''d\theta'' = \fracl{r' d\theta'}{\sqrt{1- \fracl{v^2}{c^2}}} ,
\quad
dt'' = \sqrt{1- \fracl{v^2}{c^2}}dt' -
\fracl{\fracl{v}{c^2}r' d\theta'}{\sqrt{1- \fracl{v^2}{c^2}}} \eqno{(6')}
$$
Here $V=\omega r$ is the linear speed of the rotating disk. The transformations (6') 
are written in view of the fact that the radial component of local coordinates is
perpendicular to the velocity $\vec V = [\vec \omega, \vec r]$ of the disk, 
and, according to the postulates of special relativity, does not change.
The infinitesimal  coordinate $r'd\theta'$ shrinks by the factor 
$(1- \fracl{v^2}{c^2})^{\frac{1}{2}}$ since this coordinate
is parallel to the movement direction.
The local time $dt''$ in (6') consists of two summands according to (2').

One can easily show that the ``Lorentz transformation'' for the infinitesimal coordinates
on the rotating disk has the form
$$
dr'' = dr, \quad
r''d\theta'' = \fracl{r d\theta - \omega r dt}{\sqrt{1- \fracl{\omega^2 r^2}{c^2}}},
\quad
dt'' =  \fracl{ dt - \fracl{\omega r}{c^2}r d\theta}{\sqrt{1- \fracl{\omega^2 r^2}{c^2}}}
\eqno{(7')}
$$
and the transformation (7') preserves $ds^2$:
$$
ds^2 = c^2 dt^2 - dr^2 - r^2 d\theta^2= c^2 {dt''}^2 - d{r''}^2 - {r''}^2 d{\theta''}^2
\eqno{(8')}
$$ 
The following important note concerning the transformation (7') need to be made.
The transformation (3') can be obtained from the transformation (2) via differentiation.
And  (7') is an {\em infinitesimal}  transformation connecting $dr$, $d\theta$, $dt$ with 
$dr''$, $d\theta''$, $dt''$. The corresponding equations are not completely integrable,
so it is impossible to express $r''$, $\theta''$, $t''$
in terms of $r$, $\theta$, $t$ in a
finite way!
In fact, this is a peculiar feature of Lorentz transformation from an 
inertial reference system to an accelerated one.
Transformations  of this type are usually
called local transformations, and are widely used
(see e.g. [3], [5]). We can only note that,
though there is no global connection between coordinates,
we can consider
$r''$, $\theta''$, $t''$ as ordinary coordinates
in the rotating reference system, and use them to
solve mechanical and optical problems with respect to a rotating reference system (see e.g. 
discussion on the Sagnac effect in [3]).
We suppose to construct mechanics with respect
to rotating reference system in next paper.

According to the standard rules [4],
from the space-time metric we separate out the space metric
$$
{dl''}^2 = {dr''}^2 + {r''}^2 d{\theta''}^2,
\eqno{(9')}
$$
thus on the disk we get the Euclidean geometry, contrary to the conclusions of other
researchers [4], [6] who obtained that
\begin{equation} \label{E6}
{dl''}^2 = {dr''}^2 + \fracl{{r''}^2 d{\theta''}^2}{(1- \fracl{\omega^2 {r''}^2}{c^2})} ,
\end{equation}
The difference between  (9') and (\ref{E6}) comes from the fact
that they did not take into
account the transformation of time in the rotating reference system.

The difference between our results on geometry of rotating disk and
the results of the majority of
scientists who believe that on the rotating disk
the non-Euclidean geometry arises,
leads to the difference in interpretations of effects in
rotating reference system.

From our results it follows that, according to (8'), (9'),  any experiment in a 
rotating reference system, which do not take into account
the dynamical influence
of rotation causing the centrifugal force and the Coriolis force, cannot
detect the rotation of reference system. Therefore, the transit times for
the light going in  opposite directions
around a circle $r'' = const$ coincide:
$dt''_+ = dt''_- $, hence 
\begin{equation} \label{E7}
(t''_+)_{r=r_0} = (t''_-)_{r=r_0}.
\end{equation}
Substituting (\ref{E7}) into (7'), with $r = R$, we get
\begin{equation} 
(dt_+)_{r=R} - (dt_-)_{r=R} = 2 \fracl{\omega R^2}{c^2} d\theta.
\end{equation}
Then the total time difference  with respect to the inertial reference system is
\begin{equation}  \label{E9}
\Delta t = t_+ - t_- =  \fracl{4\omega \pi}{c^2} R^2 = 2 \fracl{\omega}{c^2}S,
\end{equation}
where $S$ is the area of circle of radius $R$.

The formula (\ref{E9}) gives the exact value of difference in the light transit times, and
is called the Sagnac effect [6].

\begin{center}
References
\end{center}

1. V.Bashkov, {\em Pieces of history of Special Relativity}, Inter. conf.``Geometrization of
Physics'', Kazan, Oct. 28. -- Nov. 1, 1995, pp.62--65.

2. H.A.Lorentz, K. Verh, Akad. Wet, 1892, Bd. 1, p.74.

3. C. M\''oller {\em The Theory of Relativity}, Clarendon Press, Oxford, 1972, 182 p.

4. L.D.Landau, E.M.Lifshitz, {\em Theoretical Physics}, V.2, Field Theory, Moscow,
``Nauka'', 1973, pp.325-326.

5. R.F.Polishchuk, {\em Tetrad currents in general relativity},
Gravit. \& Cosmol., V.3(1997), No 2.(10), pp.123--129.

6. Tonnelat, M.A., {\it Les principles de la th\'eorie
\'electromagnetique et de la relativit\'e}, Paris, 1959.

\bigskip

{\bf Address:} {\it Kazan State University, Kremlevskaya, 18, Kazan: 420008, Russia}

{\bf E-mail address}: {\it Victor.Bashkov@ksu.ru, Mikhail.Malakhaltsev@ksu.ru}

\end{document}